\documentclass[prb,aps,tightenlines,twocolumn,floatfix,superscriptaddress,showpacs,preprintnumbers,citeautoscript,10pt]{revtex4-1}

\usepackage{graphicx,verbatim,multirow,dcolumn}
\usepackage{mathtools}

\usepackage{amsmath,amssymb,bm}
\usepackage{comment}
\usepackage{mwe}
\usepackage{lettrine}
\usepackage{xcolor}
\usepackage{array}
\usepackage{float}
\usepackage{tikz}
\usepackage{url}
\usepackage{soul}
\usepackage[utf8]{inputenc}
\usepackage{lipsum}
\usepackage{natbib}

\makeatletter

\newcommand{\Rmnum}[1]{\expandafter\@slowromancap\romannumeral #1@}
\makeatother

\def\<{\langle}
\def\>{\rangle}

\newcommand{\ie}[0]{\textit{i.e.}, }
\newcommand{\eg}[0]{\textit{e.g.}, }
\newcommand{\via}[0]{\textit{via} }
\newcommand{\cf}[0]{\textit{cf}.\ }

\newcommand{\DefineAuthor}[2]{%
  \expandafter\newcommand\csname #1note\endcsname[1]{%
    \textbf{\textcolor{#2}{#1: ##1}}}%
  \expandafter\newcommand\csname #1\endcsname[1]{
    \textcolor{#2}{##1}}
  \expandafter\newcommand\csname #1cancel\endcsname[1]{%
    \textcolor{#2}{\sout{##1}}}%
  \expandafter\newcommand\csname #1change\endcsname[2]{%
    \textcolor{#2}{\sout{##1} ##2}}%
  \newenvironment{#1text}{\color{#2}}{\color{black}}
}

\definecolor{dartmouthgreen}{rgb}{0.05, 0.5, 0.06}
\definecolor{cyan2}{HTML}{00F9DE}
\DefineAuthor{RAD}{red} 
\DefineAuthor{YY}{yellow} 
\DefineAuthor{GP}{cyan} 
\DefineAuthor{LK}{black} 
\DefineAuthor{OPEN}{cyan2}
\DefineAuthor{TQ}{orange} 
\DefineAuthor{AMR}{blue} 
\DefineAuthor{AMS}{magenta} 
\begin{document}
\newcommand{\tq}[1]{{\color{orange} [TQ]: #1 }}

\title{Range-Separated Hybrid Functional Pseudopotentials}

\author{
Yang Yang
}
\thanks{These authors contributed equally to this work.}
\affiliation{
Department of Chemistry and Chemical Biology, Cornell University, Ithaca, NY 14853, USA
}
\author{
Georgia Prokopiou
}
\thanks{These authors contributed equally to this work.}
\affiliation{
Department of Molecular Chemistry and Materials Science, Weizmann Institute of Science, Rehovoth 76100, Israel 
}
\author{
Tian Qiu
}
\affiliation{
Department of Chemistry, University of Pennsylvania, Philadelphia, PA 19104, USA
}
\author{
Aaron M. Schankler
}
\affiliation{
Department of Chemistry, University of Pennsylvania, Philadelphia, PA 19104, USA
}
\author{
Andrew M. Rappe
}
\affiliation{
Department of Chemistry, University of Pennsylvania, Philadelphia, PA 19104, USA
}
\author{
\\Leeor Kronik
}
\affiliation{
Department of Molecular Chemistry and Materials Science, Weizmann Institute of Science, Rehovoth 76100, Israel 
}
\author{
Robert A. DiStasio Jr.
}
\email{distasio@cornell.edu}
\affiliation{
Department of Chemistry and Chemical Biology, Cornell University, Ithaca, NY 14853, USA
}

\date{\today}

\begin{abstract}
\noindent Consistency between the exchange-correlation (xc) functional used during pseudopotential construction and planewave-based electronic structure calculations is important for an accurate and reliable description of the structure and properties of condensed-phase systems. In this work, we present a general scheme for constructing pseudopotentials with range-separated hybrid (RSH) xc functionals based on the solution of the all-electron radial integro-differential equation for a spherical atomic configuration. As proof-of-principle, we demonstrate pseudopotential construction with the PBE, PBE0, HSE, and sRSH (based on LC-$\omega$PBE0) xc functionals for a select set of atoms, and then investigate the importance of pseudopotential consistency when computing band gaps, equilibrium lattice parameters, bulk moduli, and atomization energies of several solid-state systems. In doing so, we find that pseudopotential consistency errors (PSCE) tend to be systematic and can be as large as $1.4\%$ when computing these properties.
\end{abstract}

\maketitle

\section{Introduction}\label{sec:Intro}

Density functional theory (DFT)\cite{Dreizler1990,Parr1995} has long been the computational workhorse of first-principles calculations in the fields of physics, chemistry, and materials science~\cite{Koch2001,Cramer2006,Carter2008,Sholl2009,Giustino2014,Maurer2019}. DFT is an exact theory in principle, but as it requires an exchange-correlation (xc) functional that is generally unknown, it is almost always approximate in practice. A very large number of approximate xc functionals have been suggested to date, many of which can be categorized by classes of increasing accuracy and complexity using Perdew's five-rung ``Jacob's ladder'' scheme.~\cite{Perdew_ladder} At the lower two rungs of this ladder lie the local density approximation (LDA) and generalized gradient approximation (GGA) (\eg the Perdew-Burke-Ernzerhof (PBE) functional~\cite{Perdew1996}), in which the xc term is an explicit functional of the density alone or the density and its gradient, respectively. The higher three rungs add orbital-dependent ingredients\cite{Kronik2008} of increasing sophistication, with third-rung functionals (\ie meta-GGAs) also depending on the kinetic energy density and/or Laplacian of the density, fourth-rung functionals (\ie hybrids) generally depending on the occupied orbitals, and fifth-rung functionals incorporating virtual/unoccupied orbital information. 

Hybrid xc functionals, which employ a fraction of Fock exchange (or exact exchange) as one of their ingredients~\cite{Becke1993b,Becke1993a} (\eg the hybrid PBE (PBE0) functional~\cite{perdew_rationale_1996,adamo_toward_1999}) are among the most popular fourth-rung functionals, as they often offer an excellent balance between accuracy and computational cost.~\cite{Kronik2008} Range-separated hybrids (RSH)~\cite{Savin1995,Leininger1997} are a special case of hybrid functionals, where different fractions of Fock exchange are employed at different inter-electron interaction ranges, thereby allowing for a finer balance between exchange and correlation components. Popular RSH functionals may include Fock exchange only in the short range (\eg the Heyd-Scuseria-Ernzerhof (HSE06) functional~\cite{HSE2006}, which will be referred to as HSE throughout this work), only in the long range (\eg the long-range corrected PBE (LRC-$\omega$PBE) functional~\cite{Rohrdanz2008} or the Baer-Neuhauser-Livshits (BNL) functional~\cite{Livshits2007}), or in both ranges (\eg the LRC-$\omega$PBE0 functional \cite{Rohrdanz2009}, the Cambridge-adapted Becke-3-Lee-Yang-Parr (CAM-B3LYP) functional\cite{Yanai2004}, or the $\omega$B97X functional\cite{Chai2008}). RSH functionals in which the parameters are chosen by optimal tuning, \ie by a per-system selection of parameters that satisfy physical criteria, have also been shown to be of particular use for electron and optical spectroscopy in both molecules~\cite{Kronik2012} and solids.~\cite{Kronik2018}

Solid-state DFT calculations often employ the pseudopotential (PS) method (for an overview, see: \eg Refs.\ [\onlinecite{Pickett1989, Chelikowsky1992, Singh2006, psp_review2011}] and references therein), which replaces the core electrons by an effective potential that describes their effect on the valence electrons. By removing the Coulomb singularity of the nuclear-electronic potential and eliminating the core electrons from explicit consideration, the PS method dramatically reduces the cost of solid-state DFT calculations. Clearly, a consistent level of theoretical treatment requires that the PS be generated using the same xc functional that is employed for the DFT calculation. However, virtually all PS in everyday use are generated using the lowest two rungs of Jacob's ladder, namely LDA or GGA. Given that the use of an inconsistent PS has been shown to introduce uncontrolled PS consistency errors (PSCE) even between LDA and GGA,~\cite{Fuchs1998} this is a potentially serious issue that needs to be addressed when performing more computationally intensive (and higher accuracy) DFT calculations with higher-rung xc functionals. Only recently, schemes for constructing PS for third- and fourth-rung functionals (\ie meta-GGAs and hybrids) have been reported in the literature along with an evaluation of the associated PSCE.~\cite{Yao2017,Yang2018,Bartok2019} To the best of our knowledge, PS schemes for generating fourth-rung RSH functionals have not been developed or studied to date. 

In this work, we remedy this situation. We present the general process and basic equations for the construction of RSH-type PS. Specifically, we derive the all-electron RSH radial integro-differential equation by utilizing the Slater configuration-averaging scheme~\cite{Slater,Fischer,Yang2018} in conjunction with a multipole expansion of the short-range Coulomb repulsion kernel.~\cite{Angyan2006} We then show how RSH-type PS are constructed from the all-electron orbitals and potentials. To make the entire process clear, we also show how this approach is implemented in practice (within the \texttt{OPIUM} code~\cite{opium}). Finally, we illustrate the importance of these PS by critically assessing the PSCE when computing band gaps, equilibrium lattice parameters, bulk moduli, and atomization energies of several solid-state systems. In doing so, we find that PSCE tend to be systematic and can be as large as $1.4\%$ when computing these properties using fourth-rung RSH xc functionals.

\section{Theory}\label{sec:Theory}

\subsection{Overview of RSH Functionals}

The exchange (x) energy of an RSH xc functional is split into long-range (LR) and  short-range (SR) terms,\cite{Leininger1997} often using the error function~\cite{Yanai2004} with a range-separation parameter, $\mu$. The xc energy, $E_{\rm xc}^{\rm RSH}=E_{{\rm x},\mu}^{\rm RSH}+E_{\rm c}^{\rm DFA}$, is then written as:
\begin{align}
E_{\rm{xc}}^{\rm{RSH}}&=\alpha E_{\rm{SR,}\mu}^{\rm{Fock}}+(1-\alpha)E_{\rm{SR,}\mu}^{\rm{DFA}}+(\alpha+\beta)E_{\rm{LR,}\mu}^{\rm{Fock}} \nonumber \\
&+(1-\alpha-\beta)E_{\rm{LR,}\mu}^{\rm{DFA}}+E_{\rm{c}}^{\rm{DFA}} ,
\label{eq:RSH_energy}
\end{align}
in which DFA denotes the employed density functional approximation (\eg PBE). In the general RSH scheme, $\alpha$, $\beta$, and $\mu$ are free parameters (various choices of which are discussed below), and the SR- and LR-Fock terms are given by:
\begin{align}
    E_{\rm {SR},\mu}^{\rm Fock}=-&\sum_{ij}\int d\boldsymbol{r}\,d\boldsymbol{r^\prime}\psi^{*}_{i}(\boldsymbol{r})\psi_{j}(\boldsymbol{r})\frac{\text{erfc}(\mu\left|\boldsymbol{r}-\boldsymbol{r^\prime}\right|)}{\left|\boldsymbol{r}-\boldsymbol{r^\prime}\right|}\nonumber\\
    &\times\psi^{*}_{j}(\boldsymbol{r^\prime})\psi_{i}(\boldsymbol{r^\prime})
    \label{eq:E_SRHF}
\end{align}
and
\begin{align}
    E_{\rm {LR},\mu}^{\rm Fock}=-&\sum_{ij}\int d\boldsymbol{r}\,d\boldsymbol{r^\prime}\psi^{*}_{i}(\boldsymbol{r})\psi_{j}(\boldsymbol{r})\frac{\text{erf}(\mu\left|\boldsymbol{r}-\boldsymbol{r^\prime}\right|)}{\left|\boldsymbol{r}-\boldsymbol{r^\prime}\right|}\nonumber\\
    &\times\psi^{*}_{j}(\boldsymbol{r^\prime})\psi_{i}(\boldsymbol{r^\prime}),
    \label{eq:E_LRHF}
\end{align}
where $\psi_{i}$ and $\psi_{j}$ are the occupied wavefunctions (orbitals). Note that for simplicity here and throughout, we only consider closed-shell systems, hence the $1/2$ factor is not included in Eqs.~\eqref{eq:E_SRHF} and \eqref{eq:E_LRHF}.

Like all hybrids, RSH functionals are almost invariably employed within the generalized Kohn-Sham (GKS) scheme,~\cite{Seidl1996} \ie using a non-multiplicative exchange potential related to the Fock operator. The corresponding xc potential, $\hat{V}_{\rm{xc}}^{\rm{RSH}}=\hat{V}_{\rm{x}}+V_{\rm{c}}^{\rm{DFA}}\left(\rho\left(\boldsymbol{r}\right)\right)$, is then:
\begin{align}
\hat{V}_{\rm{xc}}^{\rm{RSH}}&=\alpha \hat{V}_{\rm{SR,}\mu}^{\rm{Fock}}+(1-\alpha)V_{\rm{SR,}\mu}^{\rm{DFA}}\left(\rho\left(\boldsymbol{r}\right)\right)+(\alpha+\beta)\hat{V}_{\rm{LR,}\mu}^{\rm{Fock}} \nonumber \\
&+(1-\alpha-\beta)V_{\rm{LR,}\mu}^{\rm{DFA}}\left(\rho\left(\boldsymbol{r}\right)\right)+V_{\rm{c}}^{\rm{DFA}}\left(\rho\left(\boldsymbol{r}\right)\right),
\label{eq:RSH_potential}
\end{align}
where the ``hat'' signifies a non-multiplicative operator and $\rho(\boldsymbol{r})$ is the electron density. As an example, if we choose PBE for the DFA (\ie the use of the PBE correlation functional in conjunciton with the SR and LR versions of PBE exchange~\cite{Henderson2008,Weintraub2009}), then setting $\alpha=0.25$, $\beta=-0.25$, and $\mu=0.11$ Bohr$^{-1}$ leads to the HSE functional.~\cite{HSE2006}

\subsection{All-Electron (AE) Calculation \label{subsec:AE}}

The first step in constructing a PS is to solve the all-electron (AE) GKS equation for a specific (reference) configuration of a given atom, namely:
\begin{align}
    &\bigg[-\frac{1}{2}\nabla^2+V_{\rm ion}+V_{\rm H}
    +\hat{V}_{\rm xc} \bigg]\psi_{n l m}
    (\boldsymbol{r})\nonumber\\
    &=\epsilon_{n l m}
    \psi_{n l m}
    (\boldsymbol{r}) ,
    \label{eq:ae}
\end{align}
in which the terms in the square brackets are the electron kinetic energy operator, the nuclear-electron attraction potential, the classical electron-electron repulsion (Hartree) potential, and the xc potential, respectively. In this expression, $\psi_{nlm}(\boldsymbol{r})$ and $\epsilon_{nlm}$ are the wavefunction and eigenvalue associated with the $nlm$ quantum numbers, respectively.

The atomic wavefunctions $\psi_{nlm}(\boldsymbol{r})$ can be written as: 
\begin{align}
    \psi_{nlm}(\boldsymbol{r})=\frac{\phi_{nl}(r)}{r}Y_{lm}(\theta,\varphi)\label{eq:AEwfc}
\end{align}
in which $\phi_{nl}(r)/r$ is the (normalized) radial part of the wavefunction and $Y_{lm}(\theta,\varphi)$ is a spherical harmonic, with the aim of simplifying Eq.~(\ref{eq:ae}) into a one-dimensional ordinary differential equation for the radial function $\phi_{nl}(r)$. However, the presence of Fock exchange in $\hat{V}_{\rm xc}$ complicates this radial transformation, as the evaluation of this contribution depends on the magnetic quantum number, $m$, which reintroduces an angular dependence. To overcome this issue, we use the concept of the ``average energy of configuration" introduced by Slater~\cite{Slater} to remove the angular dependence. This so-called Slater configuration-averaging (SCA) scheme, which was used successfully for dealing with the full-range (FR) Fock exchange in the context of constructing PS for global hybrid functionals~\cite{Yang2018} (\ie hybrid functionals with $\beta=0$ in Eq.~\ref{eq:RSH_potential}), will be explained in more detail in Sec.~I.A of the \textit{Supporting Information} (SI).


For a global hybrid, in which $\hat{V}_{\rm xc} = \hat{V}_{\rm x} + V^{\rm DFA}_{\rm c}(\rho(\boldsymbol{r})) = \alpha \hat{V}^{\rm Fock}_{\rm FR} + (1 - \alpha) V^{\rm DFA}_{\rm x}(\rho(\boldsymbol{r})) + V^{\rm DFA}_{\rm c}(\rho(\boldsymbol{r}))$, this procedure leads to the following \textit{radial} expression for the (FR-)Fock contribution:
\begin{align}
\hat{V}^{\rm Fock}_{\rm FR}&\phi_{nl}(r)=\frac{1}{r}\left[\sum_{L=0}^{2l}A^{\text{FR}}_{nl,nl,L}Y^L(nl,nl;r)\phi_{nl}(r)\right.\nonumber \\
&\left.+\sum_{n^\prime l^\prime\neq nl}\sum_{L=|l-l^\prime|}^{l+l'}B^{\text{FR}}_{nl,n^\prime l^\prime,L}Y^L(n l,n^\prime l^\prime;r)\phi_{n^\prime l^\prime}(r)\right],
\label{eq:Vx_HF}
\end{align}
\noindent in which
\begin{align}
Y^L(nl,n^\prime l^\prime;r)=
\int_0^r dr^\prime \left(\frac{r^\prime}{r}\right)^L\phi^*_{n l}(r^\prime)\phi_{n^\prime l^\prime}(r^\prime) \nonumber \\
+\int_r^\infty dr^\prime \left(\frac{r}{r^\prime}\right)^{L+1}\phi^*_{n l}(r^\prime)\phi_{n^\prime l^\prime}(r^\prime) 
\label{eq:Y_L}
\end{align}
and $A^{\text{FR}}_{nl,nl,L}$ and $B^{\text{FR}}_{nl,n^\prime l^\prime,L}$ are the expansion coefficients corresponding to exchange interactions between equivalent ($nl = n^\prime l^\prime$) and non-equivalent ($nl \ne n^\prime l^\prime$) electrons, respectively; explicit expressions for both are given by Eqs.~(III.14) and (III.17) in the \textbf{SI}. In this work, we use the $Y^L$ symbol (as defined in Eq.~\eqref{eq:Y_L}) for consistency with the notation in Ref.~[\onlinecite{Fischer}], which should not be confused with the $Y_{lm}$ symbol used to denote the spherical harmonics in Eq.~\eqref{eq:AEwfc}. 

To take full advantage of these results for the FR-Fock operator in the context of RSH functionals, the terms in Eq.~\eqref{eq:RSH_potential} can be rearranged as follows: 
\begin{align}
    \hat{V}_{\rm xc}^{\rm RSH} 
    &=(\alpha+\beta)\hat{V}^\mathrm{Fock}_{\rm FR}-\beta \hat{V}^\mathrm{Fock}_{\mathrm{SR},\mu}
    +\beta V^\mathrm{DFA}_{\mathrm{SR},\mu}(\rho(\boldsymbol{r}))\nonumber\\
    &+(1-\alpha-\beta)V^\mathrm{DFA}_{\rm FR}(\rho(\boldsymbol{r}))+V^\mathrm{DFA}_{\rm c}(\rho(\boldsymbol{r})) ,
    \label{eq:Vxc_RSH}
\end{align}
such that the only term that needs to be addressed is $\hat{V}^\mathrm{Fock}_{\mathrm{SR},\mu}$, \ie the SR-Fock contribution. Here, we note that Eqs.~\eqref{eq:Vx_HF}--\eqref{eq:Y_L} are based on a multipole expansion of the standard $1/|\boldsymbol{r}-\boldsymbol{r^\prime}|$ operator found in FR-Fock exchange. To evaluate the $\hat{V}^\mathrm{Fock}_{\mathrm{SR},\mu}$ contribution in Eq.~\eqref{eq:Vxc_RSH}, we need to consider the analogous expansion for the ${\rm erfc}(\mu |\boldsymbol{r}-\boldsymbol{r^\prime}|)/|\boldsymbol{r}-\boldsymbol{r^\prime}|$  operator found in SR-Fock exchange (\cf Eq.~\eqref{eq:E_SRHF}). Such an expansion has been studied in detail by \'{A}ngy\'{a}n \textit{et al.}~\cite{Angyan2006} (and used successfully by Lehtola~\cite{Susi2020} for calculating fractionally occupied atoms) and is given as:
\begin{align}
\frac{\text{erfc}(\mu\left|\boldsymbol{r}-\boldsymbol{r^\prime}\right|)}{\left|\boldsymbol{r}-\boldsymbol{r^\prime}\right|}=\sum_{L=0}^\infty\mathcal{F}_L(r,r^\prime,\mu)P_L(\cos\gamma) ,
\label{eq:srcexp}
\end{align}
in which $P_L(\cos\gamma)$ is an $L$-th order Legendre polynomial, $\gamma$ is the angle between $\boldsymbol{r}$ and $\boldsymbol{r^\prime}$, and $\mathcal{F}_L(r,r^\prime,\mu)=\mu\Phi_L(\mu r_<,\mu r_>)$ are radial expansion functions (wherein $r_<=\min{(r,r^\prime)}$ and $r_>=\max{(r,r^\prime)}$); see Sec.~II.B in the \textbf{SI} for more details. 

With Eq.~\eqref{eq:srcexp} in hand, the SCA scheme for the SR-Fock contribution can proceed in the same manner as above, following the mathematical derivation provided in Sec~II.B of the \textbf{SI}. Following this derivation, we obtain analogous expressions to those given in Eqs.~\eqref{eq:Vx_HF}--\eqref{eq:Y_L}, namely:
\begin{align}
\hat{V}^{\rm Fock}_{\rm SR,\mu}&\phi_{nl}(r)=\frac{1}{r}\left[\sum_{L=0}^{2l}A^{\text{SR}}_{nl,nl,L}\mathcal{Z}^L_\mu(nl,nl;r)\phi_{nl}(r)\right.\nonumber \\
&\left.+\sum_{n^\prime l^\prime\neq nl}\sum_{L=|l-l^\prime|}^{l+l'}B^{\text{SR}}_{nl,n^\prime l^\prime,L}\mathcal{Z}^L_\mu(n l,n^\prime l^\prime;r)\phi_{n^\prime l^\prime}(r)\right]
\label{eq:Vx_HF_SR}
\end{align}
and
\begin{align}
    \mathcal{Z}_\mu^L(n l,n^\prime l^\prime;r)&=\mu r\int_0^r dr^\prime \Phi_L(\mu r^\prime,\mu r)\phi_{nl}^*(r^\prime)\phi_{n^\prime l^\prime}(r^\prime)\nonumber\\
    &+\mu r\int_r^\infty dr^\prime \Phi_L(\mu r,\mu r^{\prime})\phi_{nl}^*(r^\prime)\phi_{n^\prime l^\prime}(r^\prime)
\end{align}
in which $A^{\text{SR}}_{nl,nl,L}$ and $B^{\text{SR}}_{nl,n^\prime l^\prime,L}$ are the corresponding SR expansion coefficients (see Eqs.~(III.14) and (III.17) in the \textbf{SI}).

With these expressions in hand (see Sec.~III.A in the \textbf{SI} for details), Eq.~\eqref{eq:ae} can now be transformed into the following Slater-averaged all-electron RSH radial integro-differential equation:
\begin{align}
    \frac{d^2 \phi_{nl}(r)}{dr^2} &= \frac{2}{r}\bigg[\frac{l(l+1)}{2r}-Z+
    Y_\mu(nl;r) \nonumber \\ 
    &+(1-\alpha-\beta)rV^{\rm{DFA}}_{\rm FR}(r)\nonumber\\
    &+\beta rV^{\rm{DFA}}_{\rm SR}(r)+rV_{\rm{c}}^{\rm{DFA}}(r)\bigg]\phi_{nl}(r)\nonumber\\
    &+\frac{2}{r}\bigg[(\alpha+\beta)X_{\rm{FR}}(nl;r)-\beta X_{\rm{SR}}(nl;r)\bigg]\nonumber\\
    &+\epsilon_{nl}\phi_{nl}(r) .
    \label{eq:AE}
\end{align}
In this expression, $Y_\mu(nl;r)$ contains the Hartree contribution as well as the equivalent-electron ($nl=n'l'$) exchange contributions in the FR- and SR-Fock terms (\cf Eqs.~\eqref{eq:Vx_HF} and \eqref{eq:Vx_HF_SR}), namely:
\begin{align}
    Y_\mu(nl;r)&=\sum_{n^\prime l^\prime}\sum_{L=|l-l^\prime|}^{l+l^\prime}A^{\text{H}}_{nl,n^\prime l^\prime,L}Y^L(n^\prime l^\prime,n^\prime l^\prime;r)\nonumber\\
    &+\left(\alpha+\beta\right)\sum_{L=0}^{2l}A^{\text{FR}}_{nl,nl,L}Y^L(nl,nl;r)\nonumber\\
    &-\beta\sum_{L=0}^{2l}A^{\text{SR}}_{nl,nl,L}\mathcal{Z}^L_\mu(nl,nl;r) ,
    \label{eq:ynl}
\end{align}
in which $A^{\text{H}}_{nl,n^\prime l^\prime,L}$ are the needed expansion coefficients given by Eq.~(III.13) in the \textbf{SI}. 
In Eq.~\eqref{eq:AE}, $X_\text{FR}(nl;r)$ and $X_\text{SR}(nl;r)$ are the non-equivalent-electron ($nl \neq n'l'$) exchange contributions in the FR- and SR-Fock terms (\cf Eqs.~\eqref{eq:Vx_HF} and \eqref{eq:Vx_HF_SR}):
\begin{align}
X_\text{FR}(nl;r)=\sum_{n^\prime l^\prime\neq nl}\sum_{L=|l-l^\prime|}^{l+l'}B^{\text{FR}}_{nl,n^\prime l^\prime,L}Y^L(n l,n^\prime l^\prime;r)\phi_{n^\prime l^\prime}(r)
    \label{eq:X_FR}
\end{align}
and
\begin{align}
X_\text{SR}(nl;r)=\sum_{n^\prime l^\prime\neq nl}\sum_{L=|l-l^\prime|}^{l+l'}B^{\text{SR}}_{nl,n^\prime l^\prime,L}\mathcal{Z}_\mu^L(n l,n^\prime l^\prime;r)\phi_{n^\prime l^\prime}(r) .
    \label{eq:X_SR}
\end{align}
Detailed expressions for the remaining SR-DFA exchange potentials ($V^{\rm{DFA}}_{\rm SR}(r)$) are provided in Sec.~III.B in the \textbf{SI}.

\subsection{Pseudopotential (PS) Generation}

In this work, we construct RSH-type PS according to the norm-conserving PS method~\cite{hsc1979}. With the set of AE radial functions, $\{ \phi_{nl}(r) \}$, for a given atom and selected reference configuration in hand (obtained by finding the solutions to Eq.~\eqref{eq:AE}), this procedure starts by generating a corresponding set of radial pseudo-functions, $\{ \widetilde{\phi}_{nl}(r) \}$, for the valence electrons. Each $\widetilde{\phi}_{nl}(r)$ is constructed to be nodeless and slowly varying for $r < r_{\rm c}$ (a user-defined cutoff radius) and $\widetilde{\phi}_{nl}(r) = \phi_{nl}(r)$ for $r \ge r_{\rm c}$, subject to the constraint that the norm of $\widetilde{\phi}_{nl}(r)$ equals that of $\phi_{nl}(r)$. The procedure to generate a valid and smooth radial pseudo-function is non-unique and several procedures, \eg those described in Refs.\ [\onlinecite{rrkj1990,Troullier1991,Garrity2014,vanSetten2018}] are widely used today. In this work, we use the optimized PS approach of Rappe-Rabe-Kaxiras-Joannopoulos (RRKJ),\cite{rrkj1990} but all considerations outlined herein are applicable to other choices as well. 

Once the $\widetilde{\phi}_{nl}(r)$ are constructed, the so-called unscreened semi-local (SL) term in the PS, $V_l^{\rm SL} (r)$, is obtained by inverting Eq.~(\ref{eq:AE}) and then subtracting the Hartree and xc contributions corresponding to the valence electrons.\cite{TrailNeeds2005} $V_l^{\rm SL} (r)$ generated from functionals that include non-local Fock exchange (\eg global hybrids and RSH functionals) will exhibit a ``non-Coulombic tail" arising from this unscreening procedure\cite{Engel2001}; in this work, this issue is remedied by the smoothing procedure suggested by Trail and Needs,\cite{TrailNeeds2005,Rappe2008} which (following the work of Engel \textit{et al.}~\cite{Engel2001} for optimized effective potentials) unscreens the pseudopotential in a way that slightly sacrifices perfect norm conservation but restores the correct asymptotic potential. To improve transferability, we also employed the designed-non-local (DNL) method,\cite{Rappe1999,Grinberg2001,Grinberg2001_2} which adds an auxiliary function, $A(r)$ (the so-called local augmentation operator), to the Kleinman-Bylander (KB) separable form for the PS:~\cite{KleinmanBylander1982}
\begin{align}
    V^{\rm PS}(\boldsymbol{r})&=V^{\rm loc}(r)+A(r) \nonumber \\ 
    &+ \sum_{lm} \frac{ \left. \left| \Delta V_l(r) \, \widetilde{\psi}_{nlm}(\boldsymbol{r}) \right. \right> \left< \left. \widetilde{\psi}_{nlm}(\boldsymbol{r})\Delta V_l(r) \right| \right.}{\left< \left. \widetilde{\psi}_{nlm}(\boldsymbol{r}) \right| \Delta V_l(r) \left| \widetilde{\psi}_{nlm}(\boldsymbol{r}) \right. \right>} .
    \label{eq:box1}
\end{align}
In this expression, $V^{\rm loc}(r)$ is the local component of the PS, which was set equal to $V^{\rm SL}_{0}(r)$ throughout this work (\ie the default choice in \texttt{OPIUM}),
\begin{align}
    \Delta V_l(r) &\equiv V^{\rm SL}_{l}(r)-V^{\rm loc}(r)-A(r) ,
    \label{eq:box}
\end{align}
and $\widetilde{\psi}_{nlm}(\boldsymbol{r}) = \left[ \widetilde{\phi}_{nl}(r)/r \right] Y_{lm}(\theta,\varphi)$ is the pseudo-wavefunction generated for the reference atomic configuration (\cf Eq.~(\ref{eq:AEwfc})). For $A(r)=0$, we note that Eq.~\eqref{eq:box1} simplifies to the standard KB form.~\cite{KleinmanBylander1982} As mentioned above, $A(r)$ can be optimized to improve transferability; see below and Sec.~IV.A in the \textbf{SI} for more details.

To assess the transferability of $V^{\rm PS}(\boldsymbol{r})$, an AE calculation on a test (\ie non-reference) atomic configuration is performed (\via Eq.~\eqref{eq:AE}) to obtain $\{ \phi_{nl}^{\rm test}(r) \}$ and $\{ \epsilon_{nl}^{\rm test} \}$. These quantities will be compared to the corresponding radial pseudo-functions and eigenvalues, $\widetilde{\phi}_{nl}^{\rm test}(r)$ and $\widetilde{\epsilon}_{nl}^{\rm \, test}$, of the test configuration. To enable this comparison, $V_{\rm ion}\left(\rho\left(\boldsymbol{r}\right)\right)$ in Eq.~\eqref{eq:ae} is replaced with $V^{\rm PS}(\boldsymbol{r})$ from Eq.~\eqref{eq:box1} to yield the following expression for $\widetilde{\psi}_{nlm}^{\rm test}(\boldsymbol{r})$ and $\widetilde{\epsilon}_{nlm}^{\rm \, test}$: 
\begin{align}
    &\bigg[-\frac{1}{2}\nabla^2+V^{\rm PS}(\boldsymbol{r})+V_{\rm H}(\widetilde{\rho}^{\rm \, test}(\boldsymbol{r}))
    +\hat{V}_{\rm xc} \bigg] \widetilde{\psi}^{\rm test}_{nlm}
    (\boldsymbol{r})\nonumber \\ &
    =\widetilde{\epsilon}^{\rm \, test}_{n l m}
    \widetilde{\psi}^{\rm test}_{n l m}
    (\boldsymbol{r}) ,
    \label{eq:psae}
\end{align}
in which $\widetilde{\rho}^{\rm \, test}(\boldsymbol{r})$ is the electron density formed from the valence pseudo-wavefunctions of the test configuration. Following the SCA procedure described above in Sec.~\ref{subsec:AE}, we arrive at the needed RSH radial integro-differential equation (\cf Eq.~\eqref{eq:AE}) for $\widetilde{\phi}_{nl}^{\rm test}(r)$:
\begin{align}
    \frac{d^2\widetilde{\phi}_{nl}^{\rm test}(r)}{dr^2} &= \frac{2}{r}\bigg[\frac{l(l+1)}{2r}-r\left[V^{\rm loc}(r)+A(r)\right]+Y_\mu(nl;r) \nonumber\\
    &+(1-\alpha-\beta)rV^{\rm{DFA}}_{\rm FR}(r) \nonumber \\
    &+\beta rV^{\rm{DFA}}_{\rm SR}(r)+rV_{\rm{c}}^{\rm{DFA}}(r)\bigg]\widetilde{\phi}_{nl}^{\rm test}(r) \nonumber \\
    &+\frac{2}{r}\bigg[(\alpha+\beta)X_{\rm{FR}}(nl;r)-\beta X_{\rm{SR}}(nl;r)\bigg] \nonumber \\
    &-2\frac{\int_0^\infty dr \, \widetilde{\phi}_{nl}(r)\Delta V_l(r)\widetilde{\phi}_{nl}^{\rm test}(r)}{\int_0^\infty dr \, \widetilde{\phi}_{nl}(r)\Delta V_l(r)\widetilde{\phi}_{nl}(r)}\Delta V_l(r)\widetilde{\phi}_{nl}(r) \nonumber \\
    &+\widetilde{\epsilon}_{nl}^{\rm \, test}\widetilde{\phi}^{\rm test}_{nl}(r) ,
    \label{eq:nl}
\end{align}
in which $\widetilde{\phi}_{nl}(r)$ corresponds to the radial pseudo-functions of the \textit{reference} atomic configuration. Here, we note that the angular part of Eq.~\eqref{eq:box1} has been integrated out during the SCA procedure, and the Hartree and xc terms (\ie $Y_\mu$, $V_{\rm FR}^{\rm DFA}$, $V_{\rm SR}^{\rm DFA}$, $V_{\rm c}^{\rm DFA}$, $X_{\rm FR}$, and $X_{\rm SR}$) are computed using the valence radial pseudo-functions from the test configuration.\cite{Rappe1999,Grinberg2001,Grinberg2001_2} We note in passing that Eq.~\eqref{eq:nl} for $\widetilde{\phi}^{\rm test}_{nl}=\widetilde{\phi}_{nl}$ (\ie the trivial case in which the test configuration is the reference configuration) simplifies to an integro-differential equation for $\widetilde{\phi}_{nl}$ (\via algebraic manipulations of Eq.~\eqref{eq:nl}, Eq.~\eqref{eq:box}, and the definition of $V_l^{\rm SL}(r)$). In the standard case where $\widetilde{\phi}^{\rm test}_{nl} \neq \widetilde{\phi}_{nl}$, $A(r)$ will influence the generated $\widetilde{\phi}^{\rm test}_{nl}$ and $\widetilde{\epsilon}^{\rm \, test}_{nl}$, and can therefore be used to improve transferability.~\cite{Rappe1999,Grinberg2001,Grinberg2001_2}

\subsection{Implementation}

To enable AE and PS calculations involving RSH functionals, the following changes were made to \texttt{OPIUM} (version 4.1)\cite{Yang2018}:
\begin{enumerate}
    \item \'{A}ngy\'{a}n \textit{et al.}~\cite{Angyan2006} suggested a non-separable analytical expansion (NSAE) as well as a separable series expansion (SSE) for $\mathcal{F}_L$, the radial expansion functions in Eq.~\eqref{eq:srcexp}. However, Yang \textit{et al.}~\cite{Yang2021} have found that the accuracy of the SSE deteriorates with increasing angular momentum and distance (\ie for large $r_{>}$) due to numerical precision issues. Hence, higher-precision (\ie beyond customary double precision) data storage and evaluation would be needed to obtain an accurate representation for $\mathcal{F}_L$ using the SSE. In the same breath, the NSAE suffers from numerical stability issues at short distances (\ie for small $r_{>}$).~\cite{Angyan2006} In this work, we compensate for both of these shortcomings by using the NSAE for large $r_{>}$ and the SSE for small $r_{>}$ (see Sec.~II.B in the \textbf{SI} and Ref.~[\onlinecite{Yang2021}] for more details).
    
    \item For the $V^{\rm{DFA}}_{\rm SR}(r)$ term in Eq.~\eqref{eq:AE}, we have implemented the $\omega$PBE exchange functional in \texttt{OPIUM} (see Refs.~[\onlinecite{Henderson2008,Weintraub2009}] and Sec.~III.B in the \textbf{SI}). This enables one to perform AE and PS calculations in \texttt{OPIUM} for $\omega$PBE-based RSH functionals (\eg the HSE and screened-RSH (sRSH)\cite{Kronik2018} functionals). 
    
    \item We have also enabled AE and PS calculations in \texttt{OPIUM} for Yukawa-based variants of these RSH functionals (see Refs.~[\onlinecite{Yukawa1935,Akinaga2008,Fabien2011,Seth2012}] and Sec.~III.B in the \textbf{SI} for details).
    
    \item We have also extended the DNL method in \texttt{OPIUM} to PS calculations involving Fock exchange (\ie Eq.~\eqref{eq:nl} for HF, global hybrids, and RSH functionals).
\end{enumerate}

\section{Results and Discussion}\label{sec:RandD}

\subsection{PS Construction}

\begin{figure*}[t!]
\begin{center}
\includegraphics[width=\textwidth]{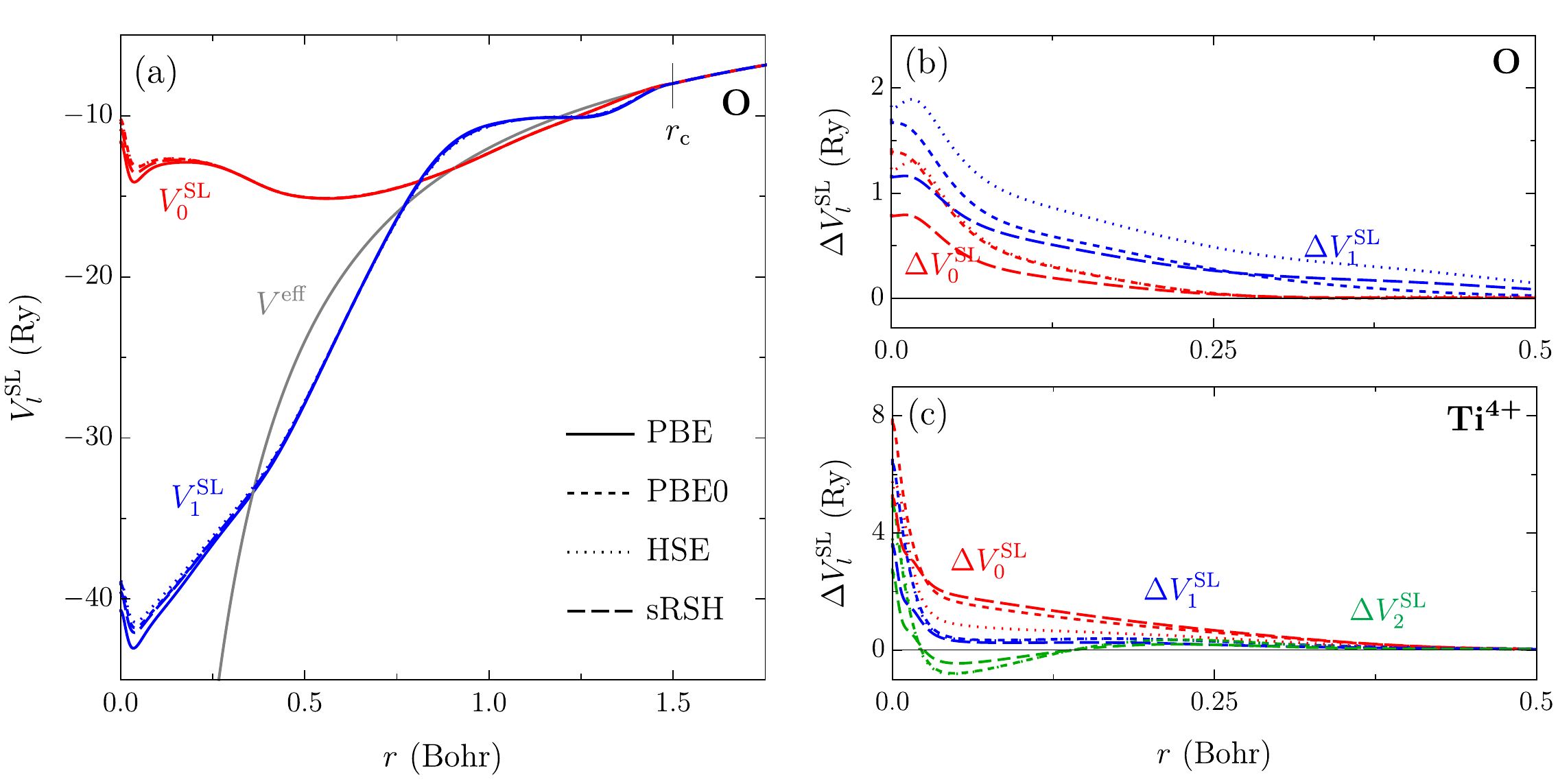}
\end{center}
\caption{\textbf{(a)} Plot of the unscreened semi-local PS term, $V_l^{\rm SL}$ ($l=0,1$) versus $r$, for the O atom generated with the PBE, PBE0, HSE, and sRSH DFAs. The effective external potential, $V^{\rm eff}(r) = -2Z_{\rm eff}/r \equiv -2(Z-Z_{\rm core})/r$, and cutoff radius, $r_c$, are also shown. For clarity,  plots of $\Delta V_l^{\rm SL} \equiv V_l^{\rm SL,DFA} - V_l^{\rm SL,PBE}$ for the PBE0, HSE, and sRSH DFAs are also depicted for \textbf{(b)} O ($l=0,1$) and \textbf{(c)} Ti$^{4+}$ ($l=0,1,2$). 
}
\label{fig:1}
\end{figure*}
We generated PS in \texttt{OPIUM} for N, O, Mg$^{2+}$, Al$^+$, P, and Ti$^{4+}$ with the PBE, PBE0, HSE, and sRSH~\cite{Kronik2016,Kronik2018} DFAs (see Sec.~IV.A in the \textbf{SI} for the parameters used for the PS construction and Sec.~IV.B in the \textbf{SI} for the sRSH parameters). Note that all RSH results reported in this work were generated using the error-function kernel in Eq.~\eqref{eq:srcexp}. In Fig.~\ref{fig:1}(a), we plot $V_l^{\rm SL}$, the unscreened semi-local PS term, for $l=0,1$ for the O atom generated with these four DFAs. In general, the PS corresponding to these four DFAs are quite similar and display more pronounced differences near the core ($r \lesssim 0.5$~Bohr). For this reason, we also plotted $\Delta V_l^{\rm SL} \equiv V_l^{\rm SL,DFA} - V_l^{\rm SL,PBE}$ for the PBE0, HSE, and sRSH DFAs for O ($l=0,1$) and Ti$^{4+}$ ($l=0,1,2$) in Fig.~\ref{fig:1}(b) and Fig.~\ref{fig:1}(c), respectively. While there are few general discernible trends that hold for both O and Ti$^{4+}$, we find that the largest differences among these PS are at $r = 0$ and range from $1.3\mathrm{-}2.2$~Ry for O and $2.8\mathrm{-}7.9$~Ry for Ti$^{4+}$.

\begin{figure}[t!]
\begin{center}
\includegraphics[width=\linewidth]{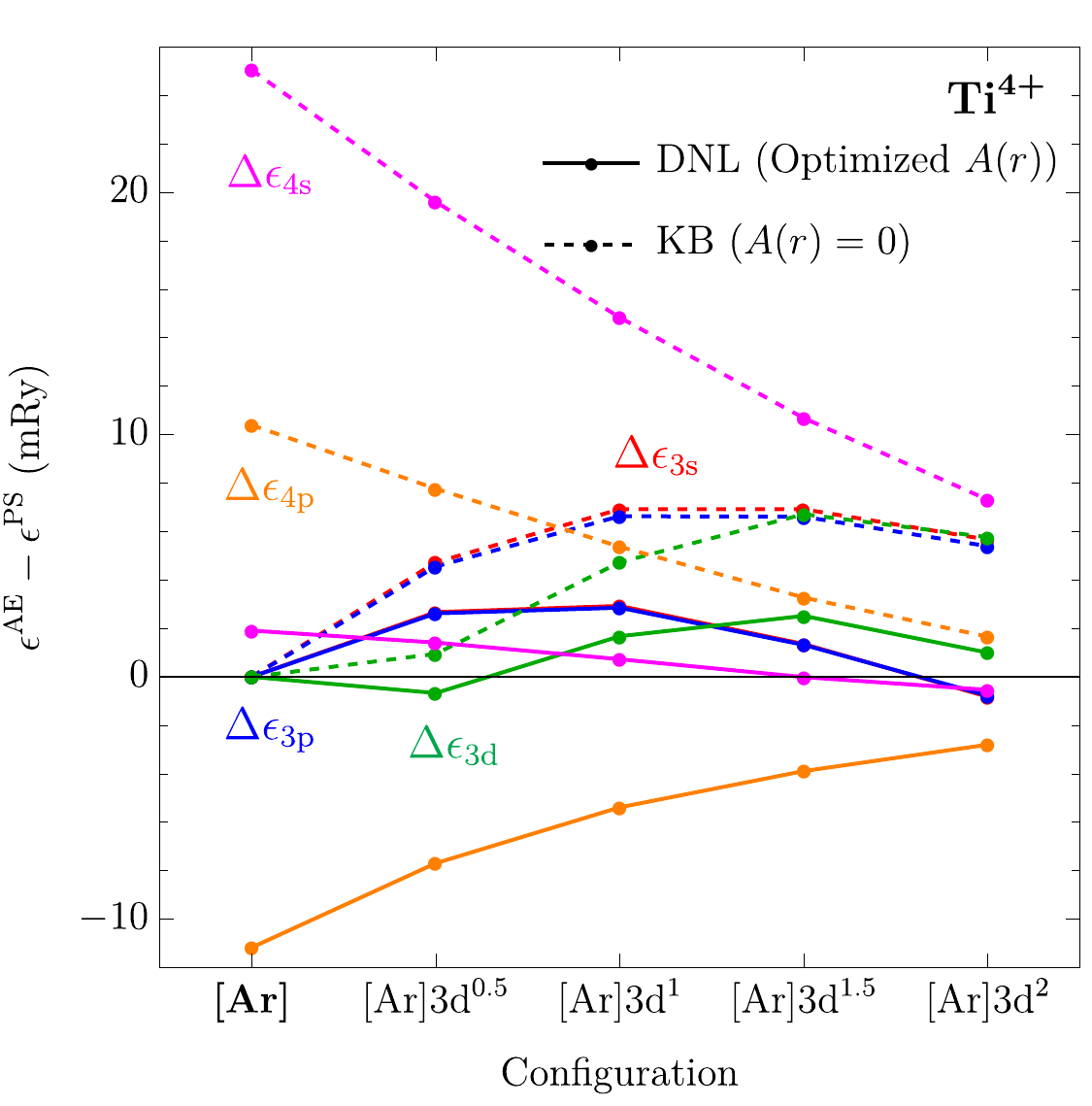}
\end{center}
\caption{Eigenvalue differences, $\Delta \epsilon \equiv \epsilon^{\rm AE} - \epsilon^{\rm PS}$, of the 3s, 3p, 3d, 4s, and 4p orbitals for different configurations of Ti$^{p+}$ ($2 \le p \le 4$; reference configuration: Ti$^{4+}$ = [Ar]) calculated using the DNL and standard KB approaches at the HSE level.
}
\label{fig:2}
\end{figure}
As mentioned above, we employed the DNL approach to improve the transferability of the PS generated in this work. This was done by optimizing $A(r)$ in Eqs.~\eqref{eq:box1}--\eqref{eq:box} to minimize the (magnitudes of the) eigenvalue differences, $\Delta \epsilon \equiv \epsilon^{\rm AE} - \epsilon^{\rm PS}$, for a select set of orbitals and configurations of a given atom. As an illustration of this approach, we optimized $A(r)$ for the 3s, 3p, 3d, 4s, and 4p orbitals across five different configurations of Ti$^{p+}$, ranging from Ti$^{4+}$ = [Ar] (reference configuration) to Ti$^{2+}$ = [Ar]3d$^{2}$. In Fig.~\ref{fig:2}, we plot $\Delta \epsilon$ for these orbitals and configurations calculated using the DNL (optimized $A(r)$) and KB ($A(r) = 0$) approaches at the HSE level. As expected, $\Delta \epsilon = 0$ for \textit{any} $A(r)$ when considering the orbitals used during PS construction (\ie 3s, 3p, 3d) of the reference Ti$^{4+}$ configuration (see discussion below Eq.~\eqref{eq:nl}). When compared to the standard KB approach, we find that the DNL optimization results in markedly improved $\Delta \epsilon_{\rm 3s}$, $\Delta \epsilon_{\rm 3p}$, and $\Delta \epsilon_{\rm 3d}$ values for the remaining four test configurations, which is strongly indicative of improved PS transferability. For the outer unoccupied 4s orbital, which was included in the optimization of $A(r)$ (but not during PS construction), we also find a significant reduction in $\Delta \epsilon$ for all configurations. In the reference configuration alone, $\Delta \epsilon_{\rm 4s}$ decreases from $\approx 25$~mRy (KB) to $\approx 2$~mRy (DNL); even though this involves the reference configuration only, this ten-fold error reduction is another criterion for demonstrating PS transferability. For the higher-lying 4p orbital, the DNL and KB approaches yielded very similar results.

\subsection{PS Consistency Errors (PSCE)}

Naturally, the next question to ask is the degree to which the differences among the PS generated with the PBE, PBE0, HSE, and sRSH DFAs affect computed solid-state properties. To answer this question, we first computed the band gaps of MgO, AlN, TiO$_2$, and AlP with PBE, PBE0, HSE and sRSH using an in-house  modified version of \texttt{Quantum ESPRESSO v6.6}\cite{QE2} (see Sec.~IV.C in the \textbf{SI} for computational details). For each of the hybrid and RSH DFAs, we performed two calculations to quantify the PSCE in the band gap, \ie the relative difference in this solid-state property when computed using an inconsistent DFA/PS combination (\eg a solid-state calculation employing PBE0 as the DFA in conjunction with a PBE PS) versus a consistent DFA/PS combination (\eg a solid-state calculation employing PBE0 as the DFA in conjunction with a PBE0 PS). From the results shown in Table~\ref{tab:gap}, we find that the PSCE grows with the magnitude of the band gap. As such, larger band-gap materials (\eg MgO and AlN) seem to be more sensitive to PS consistency (with average PSCEs of $-1.3\%$ and $-0.8\%$, respectively), while smaller band-gap systems (\eg TiO$_2$ and AlP) are less affected by PS consistency (with average PSCEs of $-0.1\%$ and $0.3\%$, respectively). For MgO, AlN, and TiO$_2$, the PSCEs tend to be systematically negative (\ie the use of consistent DFA/PS combinations tends to yield larger band gaps), while we see the opposite trend in AlP; hence PSCE can be an uncontrolled error. We also note in passing that the magnitude of the PSCE does not seem to depend strongly on the choice of hybrid or RSH DFA.

We also studied the PSCE when computing the equilibrium lattice parameters ($a_0$), bulk moduli ($k_0$), and atomization energies ($\Delta E$) of MgO and AlP, which are the simplest (cubic) systems in Table~\ref{tab:gap} (see Sec.~IV.D in the \textbf{SI} for computational details). For the equilibrium lattice parameters, we find that this property is least sensitive to DFA choice and PS consistency. While the computed atomization energies were not very sensitive to the DFA choice, we did find that this quantity has an average (maximum) PSCE of $-0.6\%$ or $24$~meV/atom ($-0.7\%$ or $28$~meV/atom for AlP). For the bulk moduli, we find that the average (maximum) PSCE in this response property are similar at $-0.5\%$ or $0.5$~GPa ($-0.8\%$ or $0.6$~GPa for AlP). We also note that the PSCEs when computing $\Delta E$ and $k_0$ were negative in all cases considered here, \ie the use of consistent DFA/PS combinations tends to yield larger values for these solid-state properties.

\begin{table}[t]
\caption{\label{tab:gap}
Band gaps (eV) of select materials computed using consistent and inconsistent DFA/PS combinations. PS consistency errors (PSCE) were computed as relative band gap differences between inconsistent and consistent DFA/PS combinations (\ie PBE0/PBE versus PBE0/PBE0).
}
\begin{ruledtabular}
\begin{tabular}{c | c c c c}
&
MgO &
AlN &
TiO$_2$ &
AlP \\
DFA/PS & ($\Gamma{\to}\Gamma$)& ($\Gamma{\to}\Gamma$) & ($\Gamma{\to}\Gamma$)& ($\Gamma{\to}X$)
\\
\hline
\\[-6pt]
PBE/PBE&\textcolor{black}{$\ 4.55$}&$\ 4.31$ &\textcolor{black}{$\ 1.77$}&$\ 1.61$\\
\\[-2pt]
PBE0/PBE &\textcolor{black}{$\ 6.94$}  & $\ 6.35$  &\textcolor{black}{$\ 3.97$}&$\ 2.97$\\
PBE0/PBE0&\textcolor{black}{$\ 7.03$}  & $\ 6.40$  &\textcolor{black}{$\ 3.98$}&$\ 2.96$\\
\multicolumn{1}{c|}{\textbf{PSCE}}&$\boldsymbol{-1.3\%}$& $\boldsymbol{-0.7\%}$ &$\ \ \boldsymbol{ -0.2\%}$&$\ \ \boldsymbol{0.1\%}$\\
\\[-2pt]
HSE/PBE&\textcolor{black}{$\ 6.23$}&$\ 5.65$ &\textcolor{black}{$\ 3.24$}&$\ 2.31$\\
HSE/HSE&\textcolor{black}{$\ 6.31$}&$\ 5.69$ &\textcolor{black}{$\ 3.24$}&$\ 2.30$\\
\multicolumn{1}{c|}{\textbf{PSCE}}&$\boldsymbol{-1.4\%}$&$\boldsymbol{-0.8\%}$ &$\ \ \boldsymbol{-0.2\%}$&$\ \ \boldsymbol{0.4\%}$\\
\\[-2pt]
sRSH/PBE&\textcolor{black}{$\ 7.90$}&$\ 6.62$ &\textcolor{black}{$\ 2.93$}&$\ 2.59$\\
sRSH/sRSH&\textcolor{black}{$\ 8.01$}&$\ 6.69$ &\textcolor{black}{$\ 2.94$}&$\ 2.58$\\
\multicolumn{1}{c|}{\textbf{PSCE}}&$\boldsymbol{-1.3\%}$&$\boldsymbol{-1.0\%}$ &$\ \  \boldsymbol{-0.1\%}$&$\ \ \boldsymbol{0.3\%}$\\
\\[-6pt]
\hline
\\[-5pt]
\multicolumn{1}{c|}{Expt.~\cite{Wing2021} } &$8.4$  &$6.5$  &$\ \ 3.0$\cite{Gap_ref1} &$2.5$ \\
\end{tabular}
\end{ruledtabular}
\end{table}

In general, our results agree with the general consensus that PSCEs tend to be smaller than the error due to the use of different DFAs.\cite{Fuchs1998,Yang2018,Borlido2020} While the PSCEs shown above are non-negligible (with an average of $0.4\%$) and can reach $-1.4\%$, this work further quantifies the errors made when using hybrid or RSH DFAs in conjunction with commonly available PBE-based PS to compute a number of solid-state properties. Here, we emphasize that PSCE is a completely avoidable error, especially when performing electronic structure calculations at the more demanding hybrid and RSH levels, and this work remedies this long-standing issue. We also note in passing that better agreement with the experimental values in Table~\ref{tab:gap} and Table~\ref{tab:A0_BM} can be obtained using a multi-projector PS, \eg the optimized norm-conserving Vanderbilt pseudopotential (ONCVP) method;\cite{Hamann2013} since this matter is not directly related to PS consistency errors, it was not pursued any further in this work.

\begin{table}[t!]
\caption{\label{tab:A0_BM}%
Equilibrium lattice parameters ($a_0$ in \AA), bulk moduli ($k_0$ in GPa), and atomization energies ($\Delta E$ in eV/atom) of select materials computed using consistent and inconsistent DFA/PS combinations. PS consistency errors (PSCE) were computed as relative property-specific differences between inconsistent and consistent DFA/PS combinations (\ie PBE0/PBE versus PBE0/PBE0).
}
\begin{ruledtabular}
\begin{tabular}{c | c c c c c c}
& 
\multicolumn{3}{c}{MgO} &
\multicolumn{3}{c}{AlP} \\
DFA/PS & $a_0$ & $k_0$ & $\Delta E$ & $a_0$ & $k_0$ & $\Delta E$
\\
\hline
\\[-6pt]
PBE/PBE& \textcolor{black}{$4.27$}&\textcolor{black}{$147.4$} & \textcolor{black}{$\ 4.74$} & $5.52$& $81.7$ &\textcolor{black}{$\ 4.12$} \\
\\[-2pt]
PBE0/PBE &\textcolor{black}{$4.22$}&\textcolor{black}{$164.5$} & \textcolor{black}{$\ 4.70$} &  $5.48$& $89.7$ &$\ 4.10$ \\
PBE0/PBE0 &\textcolor{black}{$4.22$}&\textcolor{black}{$164.9$} & \textcolor{black}{$\ 4.73$} &$5.48$& $90.4$ & \textcolor{black}{$\ 4.13$}\\
\multicolumn{1}{c|}{\textbf{PSCE}} &$\ \boldsymbol{0.0\%}$&$\ \boldsymbol{-0.2\%}$ & $\ \ \boldsymbol{-0.6\%}$ &$\ \boldsymbol{0.1\%}$& $\boldsymbol{-0.8\%}$ & $\boldsymbol{-0.7\%}$\\
\\[-2pt]
HSE/PBE &\textcolor{black}{$4.22$}&\textcolor{black}{$163.9$} & \textcolor{black}{$\ 4.72$} &$5.48$& $89.2$ & \textcolor{black}{$\ 4.11$}\\
HSE/HSE &\textcolor{black}{$4.22$}&\textcolor{black}{$164.2$} & \textcolor{black}{$\ 4.75$} &$5.48$& $89.8$ & \textcolor{black}{$\ 4.13$}\\
\multicolumn{1}{c|}{\textbf{PSCE}} &$\ \boldsymbol{0.0\%}$&$\ \boldsymbol{-0.2\%}$ & $\boldsymbol{-0.5\%}$ &$\ \boldsymbol{0.1\%}$& $\boldsymbol{-0.7\%}$ & $\boldsymbol{-0.6\%}$\\
\\
sRSH/PBE &\textcolor{black}{$4.20$}& \textcolor{black}{$170.5$}& \textcolor{black}{$\ 4.70$} &$5.49$& $89.1$ & \textcolor{black}{$\ 4.10$}\\
sRSH/sRSH &\textcolor{black}{$4.21$}&\textcolor{black}{$170.7$} & \textcolor{black}{$\ 4.72$} &$5.48$& $89.7$ & \textcolor{black}{$\ 4.13$}\\
\multicolumn{1}{c|}{\textbf{PSCE}} &$\boldsymbol{0.0\%}$&$\boldsymbol{-0.1\%}$ & $\boldsymbol{-0.4\%}$ &$\ \boldsymbol{0.1\%}$& $\boldsymbol{-0.7\%}$ & $\boldsymbol{-0.6\%}$\\
\\[-6pt]
\hline
\\[-5pt]
\multicolumn{1}{c|}{Expt.\cite{Kresse2010} }& $\ 4.19$ & $165.0$ &$\ 5.20$ & $\ 5.45$ &$86.0$&$\ 4.32$ \\
\end{tabular}
\end{ruledtabular}
\end{table}

\section{Conclusions and Future Outlook}

In this work, we presented a methodology for generating RSH-type PS and a respective implementation in \texttt{OPIUM}. With the current implementation, any RSH functional can be used for PS generation. As proof-of-principle, we generated non-local PS for several atoms with GGA, hybrid, and RSH functionals. We tested the importance of PS consistency when computing a series of solid-state properties (\eg band gaps, equilibrium lattice parameters, bulk moduli, and atomization energies) of MgO, AlN, TiO$_2$, and AlP. In doing so, our findings demonstrate that PSCEs are non-negligible (with an average of $0.4\%$) and can be as large as $-1.4\%$ when computing these common solid-state properties using inconsistent (but commonly available) PBE-based PS. We also observed that PSCEs tend to be systematic (\ie most of the PSCE values are negative). As such, this work removes the need to incur this unnecessary error when performing electronic structure calculations at the more demanding hybrid and RSH levels.

For further improvement, we plan on employing the multi-projector method,~\cite{Hamann2013} which has been shown to improve the results over standard KB design, for the hybrid and RSH PS. In addition, we believe that the importance of using a consistent functional for the electronic structure calculation and underlying PS will be even more pronounced in at least two directions, \ie response properties and heavier elements, which are both known to be more sensitive to the basis set/PS.\cite{Pickard2001,Grinberg2001} In this regard, further improvements to hybrid (and RSH) functional codes~\cite{Chawla1998,Duchemin2010,Gygi2013,Lin2016,Mandal2018,Mandal2019,Mandal2020,Ko2020,Mandal2021,Ko2021} and the extension of the RSH and hybrid PS kernels to the projector augmented wave (PAW) method\cite{PAW} and scalar/fully relativistic PS generation schemes\cite{Kleinman1980,Grinberg2000,Dolg2012} will also be needed to explore these directions.

\section{Acknowledgements}
The theoretical and computational research of TQ, AMS, and AMR was supported by the U.S.\ Department of Energy, Office of Science, Basic Energy Sciences, under Award No.\ DE-FG02-07ER46431. LK was supported by the Aryeh and Mintzi Katzman Professorial Chair and the Helen and Martin Kimmel Award for Innovative Investigation. GP and YY acknowledge support from the ``Prof.\ Rahamimoff Travel Grants for Young Scientists'' program supported by the U.S.--Israel Binational Science Foundation. This material is based upon work supported by the National Science Foundation under Grant No. CHE-1945676. RAD also gratefully acknowledges financial support from an Alfred P.\ Sloan Research Fellowship. This research used resources of the National Energy Research Scientific Computing Center, which is supported by the Office of Science of the U.S.\ Department of Energy under Contract No.\ DE-AC02-05CH11231.


%

\end{document}